\documentclass[prl,twocolumn,showpacs]{revtex4}
\usepackage[dvips]{graphicx}
\usepackage{dcolumn}
\usepackage{bm}
\usepackage{amsmath}
\usepackage{ulem}

\begin{document}
\title{Transverse Magnetic Anisotropy in Mn$_{12}$-acetate: Direct Determination by Inelastic Neutron Scattering}
\author{Roland Bircher}
\author{Gr\'{e}gory Chaboussant}
\author{Andreas Sieber}%
\author{Hans U. G\"{u}del}%
\affiliation{
Department of Chemistry and Biochemistry, University of Berne, Freiestrasse 3,
CH-3000 Berne 9, Switzerland
}%
\author{Hannu Mutka}
\affiliation{
Institut Laue-Langevin, rue Jules Horowitz 6, BP 156, 38042 Grenoble Cedex 9, France}%
\date{\today}

\begin{abstract}
A high resolution inelastic neutron scattering (INS) study of fully deuterated
Mn$_{12}$-acetate provides the most accurate spin Hamiltonian parameters for this
prototype single molecule magnet so far. The Mn$_{12}$-clusters deviate from axial
symmetry, a non-zero rhombic term in the model Hamiltonian leading to excellent
agreement with observed positions and intensities of the INS peaks. The following
parameter set provides the best agreement with the experimental data: $D=-0.0570(1)$
meV, $B_{4}^0=-2.78(7)\cdot 10^{-6}$ meV, $B_{4}^4=-3.2(6)\cdot 10^{-6}$ meV and
$\mid$\textit{E}$\mid =6.8(15)\cdot 10^{-4}$ meV. Crystal dislocations are not the
likely cause of the symmetry lowering. Rather, this study lends strong support to a
recently proposed model, which is based on the presence of several molecular isomers
with distinct spin Hamiltonian parameters.
\end{abstract}
\maketitle
Single Molecule Magnets (SMM) are crystalline compounds consisting of nominally
identical polynuclear molecular complexes of transition metal ions. They exhibit
slow magnetisation relaxation phenomena at cryogenic temperatures. Discrete steps in
the magnetisation are a typical feature directly related to quantum tunneling
processes of the magnetisation (QTM). Mn$_{12}$-acetate,
[Mn$_{12}$O$_{12}$(OAc)$_{16}$(H$_{2}$O)$_{4}$]$\cdot$2HOAc$\cdot$4H$_{2}$O, was the
first reported SMM and remains the best studied up till now. Since the pionneering
work of Sessoli \textit{et al.} \cite{Sessoli93b,Sessoli93a} $\rm Mn_{12}$-acetate
has been extensively studied by INS \cite{HanspeterPRL, Zhong99}, EPR
\cite{BarraPRB,Hill98}, NMR \cite{Furukawa00}, Raman and Infrared spectroscopy
\cite{Sushkov01}, specific heat \cite{Gomes98,Luis00}, micro-Hall probe techniques,
micro-Squid techniques \cite{Thomas96,Barbara00,Chiorescu00}, magnetization
relaxation \cite{Sessoli93b,Gomes98,Zhong00,Bokacheva00} and numerous other bulk
measurements. Identifying and quantifying the interactions leading to slow
relaxation and QTM has been the focus of several studies in very recent years
\cite{ChudnovskyPRL, CorniaPRL,delBarcoPRL,HillPRL}. In this Letter we report a high
resolution inelastic neutron scattering (INS) study on fully deuterated
Mn$_{12}$-acetate. In contrast to most other studies no external magnetic field is
involved. This leads to very accurate values of the relevant interaction parameters
responsible for QTM and allows us to unambiguously discriminate between the various
microscopic models which have been proposed to account for the observed macroscopic
behavior \cite{ChudnovskyPRL,CorniaPRL}.

[Mn$_{12}$O$_{12}$(OAc)$_{16}$(H$_{2}$O)$_{4}$]$\cdot$2HOAc $\cdot$4H$_{2}$O
crystallizes in space group I$\bar{4}$, and the
[Mn$_{12}$O$_{12}$(OAc)$_{16}$(H$_{2}$O)$_{4}$] molecule occupies a position with
S$_{4}$ point symmetry \cite{LisActaC}. The 4 H$_{2}$O and 2 HOAc (acetic acid)
solvent molecules are incorporated between the Mn$_{12}$-acetate complexes, with the
2 HOAc disordered on a fourfold position. Exchange coupling between the Mn$^{3+}$
and Mn$^{4+}$ ions within the complexes leads to a $S=10$ ground state. A strong
axial anisotropy with the S$_{4}$ axis as the easy axis splits the ground state into
sublevels from $M_{S}=\pm 10$ to $M_{S}=0$ and thus creates the energy barrier which
is responsible for the slow magnetisation relaxation phenomena at cryogenic
temperatures. The appropriate spin Hamiltonian to account for this zero field
splitting in the S$_{4}$ point group is given by \cite{SessoliAngew}:
\begin{equation}
\label{eq:Hamiltonian} \hat {H}_{axial}\, = \,D\left[ {\hat {S}_z2 -
\frac{1}{3}S\left( {S + 1} \right)} \right] + B_40 \hat {O}_40 + B_44 \hat{O}_44
\end{equation}
where $\hat{O}_{4}^{0}=35 \hat{S}_{z}^{4}-\left[ {30S(S + 1) - 25}
\right]\hat{S}_{z}^{2}-6S(S + 1)+3S^{2} \left( {S + 1} \right)^{2}$ and
$\hat{O_{4}^{4}}=\frac{1}{2} \left( \hat{S}_{+}^{4}+\hat{S}_{-}^{4}\right)$ .\\ The
first term of Eq.\ (\ref{eq:Hamiltonian}) is the leading term. Numerous parameter
sets have been proposed, with those based on EPR and INS studies, which are very
similar, usually considered the most reliable \cite{BarraPRB,HanspeterPRL}. Eq.\
(\ref{eq:Hamiltonian}) cannot account for all the observed QTM phenomena, in
particular tunneling through $M_{S}=\pm 10$ in zero field below 2 K. Additional
terms to the ones in Eq.\ (\ref{eq:Hamiltonian}) are therefore needed. A Hamiltonian
including a rhombic term of the form
\begin{equation}
\label{eq:Eterm} \hat{H}_{aniso}=\hat{H}_{axial}+E \left( \hat {S}_{x}^{2} -
\hat{S}_{y}^{2} \right)
\end{equation}
requires a deviation from the crystallographically determined S$_{4}$ molecular
symmetry. This can result from the disorder in the solvent structure, and this
possibility was examined in detail in Refs \cite{CorniaPRL,ParkPRB}. The presence of
six different geometrical molecular isomers with slightly different environments and
thus different \textit{D} and \textit{E} parameters was postulated. An alternative
model, in which the fourfold molecular symmetry is broken by crystal dislocations
was proposed in Ref.\ \cite{ChudnovskyPRL}. This model corresponds to a broad
distribution of site geometries with \textit{E} values broadly distributed around
$\mid$\textit{E}$\mid$ $=0$. Very recent EPR and QTM studies strongly favor a
discrete distribution \cite{delBarcoPRL,HillPRL}, but exact parameter values to test
the specific predictions of Refs \cite{CorniaPRL,ParkPRB} are still missing.

\begin{figure}
\includegraphics[width=84mm]{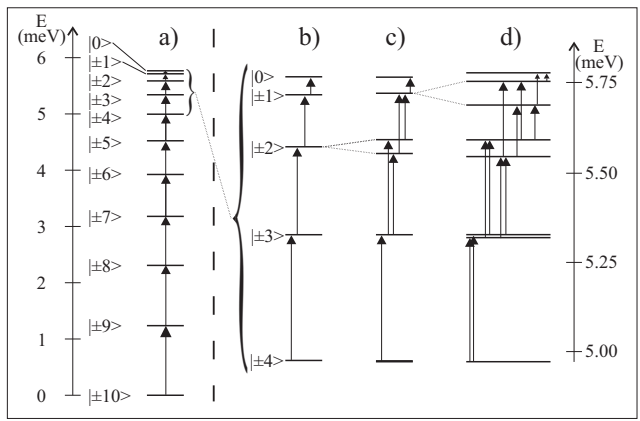}
\caption{Energy level splitting of the $S=10$ ground state and allowed (energy loss)
INS transitions. a) Total splitting, purely axial. b) Upper part of a). c) Splitting
of Eq.\ (\ref{eq:Hamiltonian}) and d) splitting of Eq.\ (\ref{eq:Eterm}) with
parameters in set I of Table \ref{table:parameters}.} \label{fig:energylevels}
\end{figure}

\begin{figure}
\includegraphics[width=80mm]{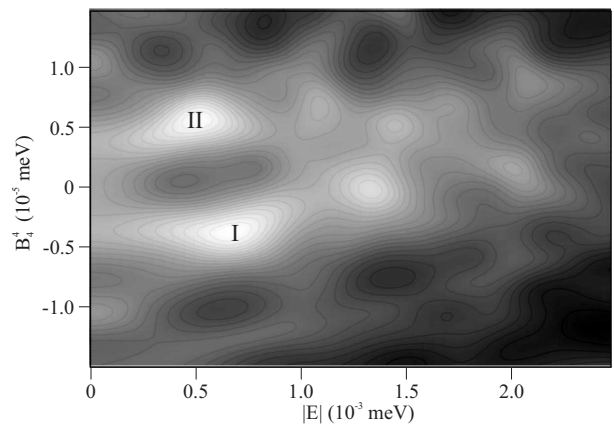}
\caption{Contour plot of the goodness of the fit of Eq.\ (\ref{eq:Eterm}) to the
experimental peak positions as a function of $B_{4}^4$ and \textit{E} with constant
values for \textit{D} and $B_{4}^0$. The two least-squares minima correspond to the
parameter sets I and II in Table \ref{table:parameters}.} \label{fig:contourplot}
\end{figure}

In an earlier INS study of partially deuterated Mn$_{12}$-acetate the data were of
sufficient quality to determine the three parameters \textit{D}, $B_{4}^0$ and
$B_{4}^4$ in Eq.\ (\ref{eq:Hamiltonian}) \cite{HanspeterPRL}. The important question
of an \textit{E} term remained open. By an upgrade, the time-of-flight instrument
IN5 at the ILL in Grenoble has gained an order of magnitude in speed, and data of
significantly higher quality than in Ref.\ \cite{HanspeterPRL} are obtained. They
allow a deeper analysis in terms of Eq.\ (\ref{eq:Eterm}) and lead to a clear
distinction between the proposed models.
\begin{figure}
\includegraphics[width=80mm]{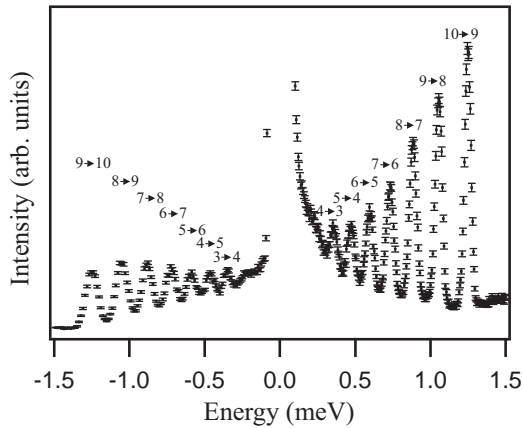}
\caption{INS spectrum measured on IN5 with an incident wavelength of $\lambda=5.9$
\AA\ at 24 K for 15 h, summed over all scattering angles. The peaks are assigned to
$\Delta M_{S}=\pm 1$ transitions, see Figure \ref{fig:energylevels}a.}
\label{fig:overview5.9}
\end{figure}
6.5 g of a fully deuterated sample of Mn$_{12}$-acetate were used in the present
study. The material was prepared as described in Ref.\ \cite{LisActaC} using
deuterated precursors and solvents. A cylindrical aluminum container of 14 mm
diameter and 55 mm length was used for the INS experiments. Data were corrected for
the background and detector efficiency using standard procedures. Figure
\ref{fig:overview5.9} shows an overview INS spectrum ($\lambda=5.9$ {\AA}) for a
temperature $T=24$ K, at which all the $M_{S}$ levels of the zero field split $S=10$
ground state have some population. The well developed and almost regular pattern of
inelastic peaks on both the energy loss and energy gain sides (positive and negative
energy transfer, respectively) are assigned to $\Delta M_{S}=\pm 1$ transitions
between adjacent ground state levels. This assignment is straightforward
\cite{HanspeterPRL}, since only $\Delta M_{S}=\pm 1$ transitions are allowed in an
axially zero field split state, see Figure \ref{fig:energylevels}a. Their relative
intensities are calculated using Ref.\ \cite{BirgenauJPCS}. The higher-order
$B_{4}^4$ and $E$ terms in Eqs.\ (\ref{eq:Hamiltonian}) and (\ref{eq:Eterm}) will
mix $M_{S}$ functions, and this becomes relevant in the following.
\begin{figure}
\includegraphics[width=80mm]{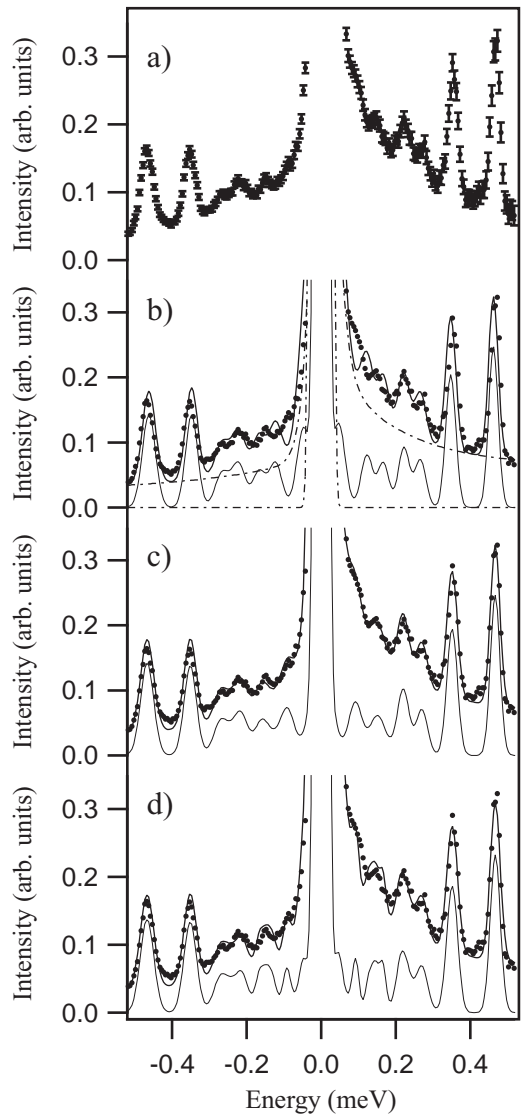}
\caption{a) INS spectrum measured with an incident wavelength of $\lambda=8$ \AA\ at
24 K for 12.5 h, summed over all scattering angles. In b) to d) the same data are
plotted without error bars including theoretical simulations. The same background,
shown as dash-dotted lines in b), was used in all the calculations. b) Eq.\
(\ref{eq:Hamiltonian}), \textit{D}, $B_{4}^{0}$ and $B_{4}^{4}$ values in set I of
Table \ref{table:parameters}. c) Eq.\ (\ref{eq:Eterm}), parameter set I in Table
\ref{table:parameters}. d) Eq.\ (\ref{eq:Eterm}), parameter set III in Table
\ref{table:parameters}.} \label{fig:experiment8A}
\end{figure}
The inner part of the INS pattern was measured at the same temperature with
increased experimental resolution ($\lambda=8$ \AA, \textit{fwhm} of the elastic
line 23 $\mu$eV). The result, displayed in Figure \ref{fig:experiment8A}a, shows a
sharpening of both the elastic and inelastic features and, in particular, some well
resolved structure below 0.3 meV on both the gain and the loss side. In this
spectral range the energy intervals and the relative intensities significantly
deviate from the regular pattern observed in Figure \ref{fig:overview5.9}. These
deviations reflect the higher order terms in the spin Hamiltonian, and they will now
be analysed. The data in Figure \ref{fig:experiment8A} are reproduced several times
without error bars to allow comparisons with different theoretical models.
\begin{table*}
\begin{tabular}{ccccccc}
 & \hspace{60pt} & \hspace{60pt}  & \textit{D}  & $B_{4}^{0}$ & $B_{4}^{4}$  & $\mid$\textit{E}$\mid$\\
 & & & (meV) & $(10^{-6}$ meV) & ($10^{-6}$ meV) & ($10^{-4}$ meV)\\
\hline \vspace{-8pt}\\
I & & & -0.0570(1) & -2.78(7) & -3.2(6) & 6.8(15) \\
II & & & -0.0570(1) & -2.78(7) & 5.1(7) & 5.3(9) \\
\hline \vspace{-8pt}\\
 & isomer & occupancy & & & & \\
 & 0 & 0.0625 & -0.0564 & -2.78 & -3.2 & 0 \\
 & 4 & 0.0625 & -0.0581 & -2.78 & -3.2 & 0 \\
III & 1 & 0.25 & -0.0564 & -2.78 & -3.2 & 6.9 \\
 & 2cis & 0.25 & -0.0573 & -2.78 & -3.2 & 0.01 \\
 & 2trans & 0.125 & -0.0573 & -2.78 & -3.2 & 13.8 \\
 & 3 & 0.25 & -0.0573 & -2.78 & -3.2 & 6.9 \\
\end{tabular}
\caption{Parameter values obtained from fits to the experimental INS peak positions.
I and II correspond to the minima in the goodness of fit plot in Figure
\ref{fig:contourplot}. Set III corresponds to the model proposed in Ref.\
\cite{CorniaPRL}, but with \textit{D} and $\mid$\textit{E}$\mid$ parameter values
from Ref.\ \cite{ParkPRB}. The numbering of isomers is the same as in Refs
\cite{CorniaPRL,ParkPRB}. The $B_{4}^{0}$ and $B_{4}^{4}$ parameters are the same as
in set I.} \label{table:parameters}
\end{table*}
The curve in Figure \ref{fig:experiment8A}b corresponds to a calculation using Eq.\
(\ref{eq:Hamiltonian}) with the \textit{D}, $B_{4}^{0}$ and $B_{4}^{4}$ parameter
values in set I of Table \ref{table:parameters}. This is the best parameter set
within the axial approximation, and it is the same within experimental error to the
one derived previously using INS \cite{HanspeterPRL}. It is evident that for energy
transfers below 0.3 meV there is poor agreement with the experimental data of the
present study. The upper part of the energy splitting pattern for this calculation
is shown in Figure \ref{fig:energylevels}c. The $B_{4}^{4}$ term splits and mixes
the $M_{S}=\pm 2$ states in first order.
Fitting our observed peak positions with the eigenvalues of Eq.\ (\ref{eq:Eterm}),
i.e. including an \textit{E} term, leads to the calculated curve in Figure
\ref{fig:experiment8A}c. The best parameters are listed in set I of Table
\ref{table:parameters}. The calculated energy pattern is shown in Figure
\ref{fig:energylevels}d. The \textit{E} term splits and mixes the $M_{S}=\pm 1$
states in first order, and this splitting is significant. This calculation shows a
remarkably good agreement with the experiment, both in terms of peak positions and
intensities. An almost equally good agreement is obtained with the parameter set II
of Table \ref{table:parameters}. The two solutions I and II correspond to negative
and positive values of the parameter $B_{4}^4$, respectively. In a purely axial
model, eigenvalues for equal positive and negative $B_{4}^4$ terms are degenerate.
Figure \ref{fig:contourplot} shows a contour plot, in which the goodness of the
least-squares fit is plotted in the parameter space spanned by $B_{4}^4$ and
$\mid$\textit{E}$\mid$. The two minima corresponding to the solutions I and II in
Table \ref{table:parameters} are very well defined.
We confidently conclude that an \textit{E} term is necessary to account for the
observed INS data. And in contrast to earlier studies we can quantify the
$\mid$\textit{E}$\mid$ parameter: $\mid$\textit{E}$\mid=6.8(15)\cdot 10^{-4}$ meV
and $\mid$\textit{E}$\mid=5.3(9)\cdot 10^{-4}$ meV for solutions I and II,
respectively.  Of the two parameter sets we give preference to set I, i.e. a
negative $B_{4}^{4}$ parameter, because our value $B_{4}^4=-3.2(6)\cdot 10^{-6}$ meV
lies close to the most recent and very accurate determination by EPR of
$\mid$\textit{B}$_{4}^{4}\mid=3.8(1)\cdot 10^{-6}$ \cite{HillPRL}.

Of the two models proposed in the literature to account for a deviation from axial
molecular symmetry we definitely rule out the dislocation model
\cite{ChudnovskyPRL}, because it corresponds to a broad distribution of sites with
\textit{E} terms centered around $\mid$\textit{E}$\mid=0$. This is the energy level
pattern in Figure \ref{fig:energylevels}c and would lead to a broadened version of
the calculated spectrum in Figure \ref{fig:experiment8A}b. The alternative model
with a set of six discrete sites for the Mn$_{12}$-acetate molecule \cite{CorniaPRL}
has recently received qualitative support from both EPR and tunneling studies
\cite{delBarcoPRL,HillPRL}. In Ref.\ \cite{CorniaPRL} the $D$ and $E$ parameters of
the six isomers were estimated on the basis of the angular overlap model. Based on
the same isomer distribution model $D$ and $E$ parameters obtained by a density
functional theory (DFT) calculation were recently reported \cite{ParkPRB}. The
distribution of $D$ and $E$ values is very similar to Ref.\ \cite{CorniaPRL}, but
the absolute values are more reliable. We calculated the energy splitting patterns
for the six isomers and the corresponding INS spectrum at 24 K, using the $D$ and
$\mid$\textit{E}$\mid$ values from Ref.\ \cite{ParkPRB} and our best $B_{4}^{0}$ and
$B_{4}^{4}$ values, see set III in Table \ref{table:parameters}. The result is shown
in Figure \ref{fig:experiment8A}d. The agreement with the experimental data is
remarkably good, considering that no adjustment of the $D$ and
$\mid$\textit{E}$\mid$ parameters was attempted. It is not \textit{a priori} clear
why the distribution of six species with populations and parameter values as given
in set III of Table \ref{table:parameters} gives a discrete INS spectrum with well
defined peaks. Inspection of the parameter set III in Table \ref{table:parameters}
reveals that the three isomers 1, 2 cis and 3 make up 75\% of all complexes in the
crystal. The isomers 1 and 3 have very similar $D$ and, more importantly in this
context, practically identical $\mid$\textit{E}$\mid$ parameters to our best
parameter set I. This, together with some near coincidences of transition energies
for the other isomers, leads to a discrete spectrum which is in good agreement with
experiment. Our study thus provides support for the model proposed in Ref.\
\cite{CorniaPRL} and refined in Ref.\ \cite{ParkPRB}.

Our spectroscopic study thus clearly reveals that the Mn$_{12}$-acetate clusters in
[Mn$_{12}$O$_{12}$(OAc)$_{16}$(H$_{2}$O)$_{4}$]$\cdot$2HOAc$\cdot$4H$_{2}$O deviate
from axial symmetry. A rhombic term in the spin Hamiltonian is essential, and all
three parameter sets in Table \ref{table:parameters} reproduce the INS spectra very
well. Earlier studies based on Landau-Zener (LZ) magnetisation relaxation
\cite{delBarcoPRL} and EPR \cite{HillPRL} measurements provided upper limits of
$8.7\cdot 10^{-4}$ meV and $12.4\cdot 10^{-4}$ meV, respectively, for the
$\mid$\textit{E}$\mid$ values in the most strongly rhombically distorted complexes
in Mn$_{12}$-acetate. Our study significantly narrows down the range of possible
$\mid$\textit{E}$\mid$ values. In contrast to LZ and EPR measurements we are not
primarily probing the complexes with the fastest relaxation or the strongest rhombic
distortion but the total of all complexes. And the
$\mid$\textit{E}$\mid=6.8(15)\cdot 10^{-4}$ meV value of set I in Table
\ref{table:parameters} represents this. If we adopt the isomer distribution model,
this value accounts for 50\% of all the complexes; for 37.5\% the
$\mid$\textit{E}$\mid$ value is at least an order of magnitude smaller, and for
12.5\% it is $13.8\cdot 10^{-4}$ meV. This latter value is in reasonable agreement
with upper limit estimates of $8.7\cdot 10^{-4}$ meV and $12.4\cdot 10^{-4}$ meV
from LZ and EPR, respectively. Interestingly, it was recently suggested, based on
EPR experiments, that in deuterated Mn$_{12}$-acetate the upper limit of
$\mid$\textit{E}$\mid$ was $24.8\cdot 10^{-4}$ meV, twice as high as for the
undeuterated material \cite{delBarcoCondmat0404390}. This value is high and outside
the range of our parameter values. Our experiment was performed on a completely
deuterated sample, and we took great care to handle it in a water free atmosphere .
In a sample which is not fully deuterated there is additional disorder in the acetic
acid and water structure, and this could possibly lead to a broader distribution of
rhombic distortions and thus $\mid$\textit{E}$\mid$ values. Rhombic distortions of
individual complexes, likely caused by disorder in the solvent structure, are
responsible for some of the observations in QTM measurements which are not
compatible with a purely axial model. The present study provides quantitative
information about the size of the rhombic parameters. This model is not able to
account for the observation in Mn$_{12}$-acetate of QTM between levels with odd
$\Delta M$ values. Possible mechanisms have been proposed in Refs
\cite{ChudnovskyPRL,CorniaPRL,Hill04_condmat}. These effects must be small and our
experiment provides no information to quantify them.

The authors acknowledge useful discussions with Oliver Waldmann. This work was
supported by the Swiss National Science Foundation and the TMR programmes Molnanomag
and Quemolna of the European Union (HPRN-CT-1999-00012 and MRTN-CT-2003-504880).


\begin{thebibliography}{}

\bibitem{Sessoli93b} R. Sessoli {\it et al}, Nature {\bf 365}, 141 (1993).
\bibitem{Sessoli93a} R. Sessoli {\it et al}, J. Amer. Chem. Soc. {\bf 115}, 1804 (1993).
\bibitem{HanspeterPRL} I. Mirebeau {\it et al}, Phys. Rev. Lett. {\bf 83}, 628 (1999).
\bibitem{Zhong99} Y. Zhong {\it et al}, J. App. Phys. {\bf 85}, 5636 (1999).
\bibitem{BarraPRB} A. L. Barra, D. Gatteschi and R. Sessoli, Phys. Rev. B {\bf 56}, 8192 (1997).
\bibitem{Hill98} S. Hill {\it et al}, Phys. Rev. Lett. {\bf 80}, 2453 (1998).
\bibitem{Furukawa00} Y. Furukawa {\it et al}, Phys. Rev. B {\bf 62}, 14246 (2000).
\bibitem{Sushkov01} A. B. Sushkov {\it et al}, Phys. Rev. B {\bf 63 }, 214408 (2001).
\bibitem{Gomes98} A. M. Gomes {\it et al}, Phys. Rev. B {\bf 57}, 5021 (1998).
\bibitem{Luis00} F. Luis {\it et al}, Phys. Rev. Lett. {\bf 85}, 4377 (2000).
\bibitem{Thomas96} L. Thomas {\it et al}, Nature {\bf 383}, 145 (1996).
\bibitem{Barbara00} B. Barbara {\it et al}, J. Phys. Soc. Jpn. {\bf 69}, Suppl. A., 383 (2000).
\bibitem{Chiorescu00} I. Chiorescu {\it et al}, Phys. Rev. Lett. {\bf 85}, 4807 (2000).
\bibitem{Zhong00} Y. Zhong {\it et al}, Phys. Rev. B {\bf 62}, 9256 (2000).
\bibitem{Bokacheva00} L. Bokacheva, A. D. Kent and M. A. Walters, Phys. Rev. Lett. {\bf 85}, 4803 (2000).
\bibitem{ChudnovskyPRL} E. M. Chudnovsky and D. A. Garanin, Phys. Rev. Lett. {\bf 87} , 187203 (2001).
\bibitem{CorniaPRL} A. Cornia {\it et al}, Phys. Rev. Lett. {\bf 89}, 257201 (2002).
\bibitem{delBarcoPRL}  E. del Barco {\it et al}, Phys. Rev. Lett. {\bf 91}, 47203 (2003).
\bibitem{HillPRL} S. Hill {\it et al}, Phys. Rev. Lett. {\bf 90}, 217204 (2003).
\bibitem{LisActaC} T. Lis, Acta Crystallog. Sec. B  {\bf 36}, 2042 (1980).
\bibitem{SessoliAngew} D. Gatteschi and R. Sessoli, Angew. Chem. Int. Ed. {\bf 42}, 268 (2003).
\bibitem{ParkPRB} K. Park {\it et al}, Phys. Rev. B {\bf 69}, 144426 (2004).
\bibitem{BirgenauJPCS} J. R. Birgenau, J. Phys. Chem. Solids, {\bf 33}, 59 (1972).
\bibitem{delBarcoCondmat0404390} E. del Barco {\it et al}, cond-mat/0404390 (unpublished).
\bibitem{Hill04_condmat} S. Hill {\it et al}, cond-mat/0401515 (unpublished).

\end{thebibliography}
\end{document}